\magnification=1200
\baselineskip=18truept


\def\draftversion{N}

\if \draftversion Y


\fi

\def\gaumeginsp{1}
\def\plbfirst{2}
\def\npblong{3}
\def\daemi{4}
\def\RDS{5}
\def\phillips{6}
\def\berry{7}
\def\BZ{8}
\def\simon{9}
\def\ball{10}
\def\WZcons{\BZ}
\def\npbold{11}
\def\fujik{12}
\def\QCDovlap{13}
\def\kap{14}
\def\nacul{15}
\def\sonoda{16}
\def\Avron{17}
\def\412{18}
\def\num412{19}
\def\FNN{20}
\def\prev2dinst{\npbold}
\def\PRL{21}
\def\NNintuit{22}
\def\pancha{23}
\def\hasen{24}
\def\minfty{25}

\line{\hfill RU--98--4}
\vskip 3cm
\centerline {\bf Geometrical aspects of chiral anomalies in the overlap.}
\vskip 1cm
\centerline{Herbert Neuberger}
\vskip 1.5cm
\centerline{\it Department of Physics and Astronomy}
\centerline{\it Rutgers University}
\centerline{\it Piscataway, NJ 08855--0849}
\vskip 2cm
\centerline{\bf Abstract}
\vskip .5cm
The set of one dimensional 
lowest energy eigenspaces used to construct the overlap
induces a two form on gauge orbit space which is the locally exact
curl of Berry's connection. If anomalies do not cancel, examples
of two dimensional closed sub-manifolds of orbit space are produced
over which the integral of the above two form does not vanish.
Based on these observations, a natural definition of covariant currents
is obtained, a simple way to calculate chiral anomalies on the lattice
is found, and indications for how to construct an ideal regularization
of chiral gauge theories are seen to emerge.

\vskip .3cm
\vfill
\eject
{\bf 1. Introduction.}
\vskip .5cm
In the continuum, on a compact Euclidean space-time manifold, chiral
anomalies can be understood and evaluated from solely geometric
considerations [\gaumeginsp]. 
On a finite lattice it would appear that these
insights have to be lost, being somehow restored 
in the continuum limit. In this paper the overlap will be shown
to provide a geometrical interpretation for chiral anomalies
directly on lattices approximating a continuum torus. With
this understanding it will become clearer why theories where
anomalies cancel between different multiplets are fundamentally
different in the overlap approach
from those where the anomalies do not cancel. This insight holds
directly on the lattice and does not appeal to continuum physics
or perturbative concepts.

The essence of the overlap [\plbfirst,\npblong,\daemi,\RDS] 
is the association of the continuum
chiral determinant,
viewed as a line bundle over the space of gauge field configurations
factored by gauge transformations, with a line bundle of ground
states of a certain bilinear fermionic Hamiltonian over the space of
all compact link variables factored by lattice gauge transformations.
(More precisely, the lattice complex 
line bundle consists of a collection of 
projections of the ground states along a fixed vector, denoted
by $|v_+>$ in the appendix.) 
This association is imperfect because the space of gauge orbits
in the continuum is disconnected while the analogous space on the lattice
is connected. This imperfection is reflected by the
need to excise from
the space of lattice gauge orbits those for which the dimensionality
of the ground state eigenspace exceeds unity because of accidental
degeneracies. The set of excised configurations has zero weight 
in the measure induced by the Haar volume element per link and per site
for link parallel transporters and for 
gauge transformations, respectively. 
The excised configurations are ``exceptional'' in the sense adopted
in early studies of gauge field topology on the
lattice [\phillips]. 
The removal of the exceptional configurations
leaves behind a space of lattice gauge orbits that
no longer is connected, but the number of components if finite.

The chiral determinant vanishes over all
connected components of the space of lattice orbits
except one. The Hamiltonians commute with fermion
number and the fermion number of $|v_+>$ is definite. 
Thus, the projection along $|v_+>$ vanishes
whenever the ground state has a fermion number different
from that of $|v_+>$. 
In any construction of chiral gauge theories 
one first focuses on the one component where the 
chiral determinant does not
vanish. This component is a connected,
continuous space of gauge orbits and this work is restricted to it. 
To define the chiral determinant for these backgrounds
a smooth section of the line
bundle of projected vacuum eigenspaces would be needed.

We shall find two kinds of obstacles to the construction
of a gauge invariant chiral determinant: the first consists of 
obstructions which need to mutually cancel and the second is a residue
remaining after the cancelation. 
Both obstacles can be understood in the
framework of Berry's phase [\berry].
The basic
object will be a two form over the 
space of gauge potentials, which
is locally exact and given by the curl
of $\Delta j$, the difference between the covariant
and the consistent currents [\BZ]. This abelian curvature
plays a central role also in other regularizations of the chiral
gauge theories [\ball], but in the overlap it 
acquires a simple geometrical
interpretation, becoming Berry's curvature for a certain
line bundle. 

As usual, instead of working
over the space of orbits we start by working over the space of gauge
connections, or link variables. Obstructions tell us 
that factoring by gauge transformations would not produce single
valued functions over the space of orbits. 
In section 2 we show that the usual definitions of 
the consistent and covariant currents [\BZ], $j^{\rm cons}$,
$j^{\rm cov}$ and $\Delta j$ are consistent with  $j^{\rm cons}$
being the variation of the lattice overlap w.r.t the gauge field
and $\Delta j$ being Berry's connection associated with a certain
set of finite vectors $|v>$ parameterized by the gauge fields [\simon].  
We then proceed to show that Berry's curvature $d\Delta j$ is
a non-perturbatively, gauge invariantly regularized version of
the antisymmetric two form $Z(\delta_1 A, \delta_2 A, A)$
of [\ball]. In sections 3 and 5 we show in dimensions 2 and 4
that the known continuum expressions for  $Z(\delta_1 A, \delta_2 A, A)$
can be determined directly from Berry ``monopole'' singularities.
In section 4 it is shown that in a certain two dimensional case,
for which anomalies cancel, the remaining 
lattice artifact terms in  $Z(\delta_1 A, \delta_2 A, A)$
can be diminished by smooth deformations of the constructions
of the $|v>$ line bundle. If anomalies do not cancel no such
deformations can affect a certain component of the total
curvature; this is the component that survives in the continuum
but it separates cleanly from the other contributions,
already on the lattice. 
In section 6 the usual Brillouin-Wigner phase choice
is shown to also possess a certain geometrical meaning.
It is suggested that this insight explains why the
BW phase convention turned out to obey several desirable
symmetries [\npblong]. 
Several conclusions and conjectures are described
in section 7. Appendix A contains a 
definition of the lattice overlap in any even dimension. 
Appendix B clarifies the compact notation of 
section 2 in the lattice context. 

Let me briefly discuss the generality of the results: 
Sections 3 through 5 deal with abelian gauge backgrounds. The
results of sections 3 and 5 extend to the non-abelian case
by embedding. Section 4 presents an explicit abelian example of
an improved choice of Hamiltonians. It demonstrates that by
identifying the ``monopole'' sources of anomalies in statistically
important backgrounds, one can reduce lattice artifacts in anomaly
free cases. While the example is abelian and two dimensional, 
the principle behind it is general and 
applies to any even dimension and any compact gauge group. 
The principle is explained  in section 2.

\vskip 1cm
\centerline{\bf 2. $\Delta j$ and Berry's phase in the overlap.}
\vskip .5cm

This section has two parts. In the first I review the concepts
of covariant and consistent currents in the continuum and introduce
a compact notation to emphasize the essentials of the structures. In
the second part I describe lattice objects
corresponding to the continuum currents of the first part.

\leftline{\sl 2a. Continuum.}

Since our independent variables are the continuum gauge fields I find it
easier to start by choosing a notation that makes this obvious:
Replace the fields $A^a_\mu (x)$ by
real coordinates $\xi_\alpha$. The index $\alpha$ 
takes all the values taken
by the triplet $(\mu ,a,x)$. An infinitesimal 
gauge transformation is parameterized by a function
of the pair $(a,x)$, 
and will be denoted by $\omega_i$, where $i$ replaces $(a,x)$.
The consistent current (to be defined below) is a functional of
the gauge field, a function of $x$ and has components labeled 
by $\mu$ and $a$. It will be denoted by $j^{\rm cons}_\alpha (\xi )$. 
Clearly, it can be viewed as a one 
form $j^{\rm cons}=j^{\rm cons}_\alpha d\xi_\alpha$
over the space coordinatized 
by the $\xi_\alpha$. Under a finite gauge transformation $g$, 
our coordinates $\xi_\alpha$ get replaced 
by $(\xi^g )_\beta ~$:
$$
(\xi^g )_\beta = h_\beta (g ) + {\cal D}^{-1}_{\alpha\beta}(g)
\xi_\alpha .\eqno{(2.1)}$$
The first term is an inhomogeneous global shift and the second
is a homogeneous linear transformation. The equation is nothing
but the usual gauge transformation. We shall only need it for
small variations $\delta\xi$, where the inhomogeneous term
drops out. We have
$$
{{\partial}\over {(\xi^g )_\alpha}} \xi_\beta = 
{\cal D}_{\alpha\beta} (g ),
\eqno{(2.2)}$$
where  ${\cal D}_{\alpha\beta} (g)$ is real. 

A gauge invariant function $I(\xi)$ obeys:
$$I(\xi ) = I(\xi^g ).\eqno{(2.3)}$$
Functions $\psi_\alpha (\xi)$ (a one form $\psi_\alpha (\xi )
d\xi_\alpha$) that transform as the 
gradient $\partial_\alpha I$ $(dI)$, are said to transform covariantly:
$$\psi_\alpha (\xi^g )=
{\cal D}_{\alpha\beta} (g) \psi_\beta (\xi ).\eqno{(2.4)}$$

The matrices  ${\cal D}_{\alpha\beta} (g)$ represent 
the action of a gauge transformation $g$ on 
the $\psi_\alpha (\xi)$ objects in a covariant manner. 

Associated with the Hamiltonian ${\cal H}^-$ (see appendix)  
is a Hilbert space of finite dimension, ${\cal V}$, 
providing a Fock representation for a set of canonical fermionic
creation and annihilation operators $a$, $a^\dagger$. 
Gauge transformations $g$ are
unitarily represented on this Hilbert space by $G(g)$ in a $\xi_\alpha$
independent way. 

Let us assume now that we have a regularization that produces a chiral
determinant $D(\xi )$. The consistent current is defined by:
$$
j_\alpha^{\rm cons} (\xi ) = \partial_\alpha \log D (\xi ).
\eqno{(2.5)}$$
Consistency [\WZcons] simply means,
$$
\partial_\beta j_\alpha^{\rm cons}=\partial_\alpha j_\beta^{\rm cons},
\eqno{(2.6)}$$
or
$$dj^{\rm cons } =0,\eqno{(2.7)}$$
as $j^{\rm cons} =d\log D(\xi )$. 

An infinitesimal gauge transformation acts on functions of $\xi_\alpha$
in a way dictated by the form of $(\xi^g )_\alpha$ (eq. (2.1)),
$$
\partial_i = (X_{i\alpha} + 
Y_{i\alpha\beta} \xi_\beta )\partial_\alpha\eqno{(2.8)}$$
with $\xi$-independent $X$ and $Y$. $Y$ comes from the linear
part of ${\cal D}$ expanded around $g\equiv 1$ and 
proportional to $\omega_i$. 
Hence ([\BZ]), 
$$
\partial_\alpha \partial_i - \partial_i \partial_\alpha = Y_{i\beta\alpha}
\partial_\beta ,\eqno{(2.9)}$$
which implies:
$$
\partial_\alpha (\partial_i \log D ) - \partial_i j_\alpha^{\rm cons}=
Y_{i\beta\alpha} j_\beta^{\rm cons}.\eqno{(2.10)}$$
The infinitesimal form of eq. (2.4) is $\partial_i \psi_\alpha
+Y_{i\beta\alpha} \psi_\beta =0 $. 
If anomalies are absent, $j_\alpha^{\rm cons}$
transforms covariantly because  $\partial_i \log D =0 $, 
the $Y$ implementing the familiar commutator.
A non-vanishing anomaly implies a noncovariant transformation law
for the consistent current in the non-singlet, nonabelian case.
Bardeen and Zumino (BZ) [\BZ] show by explicit construction that there always
exists a one form $\Delta j$, polynomial (in $\xi$), 
local in space-time, such that $ j^{\rm cons} + \Delta j$ 
transforms covariantly. 

In summary, the anomalous, non-abelian situation in the continuum
is as follows: There are two currents, $j^{\rm cons}$ and $j^{\rm cov}$.
$j^{\rm cov}$ is not the variation of any function, but is gauge
covariant. $j^{\rm cons}$ is the derivative of a function (the 
regulated chiral determinant) but is not gauge covariant. Given
$j^{\rm cons}$ we could reconstruct the regulated determinant,
but the lack of gauge covariance of $j^{\rm cons}$ makes the
reconstructed function break gauge invariance. When anomalies
cancel $j^{\rm cons}$ and $j^{\rm cov}$ are equal. In that
case the total reconstructed determinant is gauge invariant. 
So, when we go to the lattice, we should focus our attention
on $\Delta j$, the difference between $j^{\rm cons}$ and $j^{\rm cov}$.
We wish to understand on the lattice directly
why in the anomalous case it is unavoidable that $\Delta j \ne 0$. 
Then, we wish to see that when anomalies cancel it no longer is
unavoidable that $\Delta j \ne 0$. To make $\Delta j = 0$
one would need to tune the lattice overlap Hamiltonian matrices
(for definitions see Appendix A). 
I present an argument that the obstruction preventing $\Delta j = 0$
{\it on the lattice} in the anomalous case disappears if the
{\it continuum algebraic conditions} for anomaly cancelation hold. 
The argument is geometrical, goes to the heart of the matter, 
and is supported  by abelian examples. In the abelian case the 
compactness of the group is crucial.  

\leftline{\sl 2b. Lattice.}

On the lattice, in a nonperturbative framework, at a finite
cutoff, with compact gauge fields, $\Delta j$ will not be polynomial and 
one does not regularize just the {\it variation} of the
chiral determinant, as is sometimes 
done in the continuum [\ball]; the determinant
itself is regularized. 
One way or another, when defining the chiral determinant, one
always deals with a determinant line bundle over the space of $\xi$'s
(before attempting to factor out gauge transformations). 
The overlap is no exception, only now one has one dimensional
spaces that are naturally embedded in 
one common larger space. This allows
us to compare the one dimensional spaces over different 
points $\xi$. Therefore, one has a natural split of the
variation in a vector at $\xi$ induced by deforming $\xi$ 
to $\xi +\delta\xi$:
one part of the variation is the ``real'' 
change in the spaces spanned by each vector and the
other is the ``irrelevant'' change along the fibers. 

Since $<v_+|$ is taken to be $\xi$ independent (see Appendix A) 
the variation of the regulated chiral determinant is directly
related to the variation of a vector in ${\cal V}$:
$$
\delta <v_+|v_-> = <v_+|\delta v_->.\eqno{(2.11)}$$
Henceforth we shall suppress the
subscript minus and shall replace $|v_->$ by $|v>$.
(Similarly, the Hamiltonian superscript will be dropped.)
The split we described earlier is:
$$
\partial_\alpha |v>= ( \partial_\alpha |v>)_\perp +|v><v|\partial_\alpha|v>,
\eqno{(2.12)}$$
where $<v|[ ( \partial_\alpha |v>)_\perp ]=0$
since $|v>$ is normalized by definition.

This split was noted already in [\npbold], but the roles
of the two terms were misidentified, 
based on an example calculation
that turned out to be wrong, producing an 
incorrect coefficient.\footnote{${}^{f_1}$}
{Equation 3.70 in ref. [\npbold] has the wrong
overall sign and the first term in the round brackets
should be deleted.} This
error precluded further development until, about one year ago, 
S. Randjbar-Daemi and J. Strathdee [\RDS], starting from scratch arrived
at the same split, but this time correctly identified the roles of the 
pieces.\footnote{${}^{f_2}$}
{They also computed various quantities, including the
afore mentioned coefficient; their equation (24)
provides a correct replacement for equation 3.73
in [\npbold].}
The first term in equation (2.12), 
measuring the distance between the
spaces, corresponds to $j^{\rm cov}$, the covariant form of the current.
The second term gives $\Delta j$. (Obviously, the sum gives $j^{\rm cons}$.) 

Let me first show explicitly that $j^{\rm cov}$ so identified indeed
transforms covariantly. Under $\xi \rightarrow \xi^g$, 
a replacement of the parameters
in $\partial_\alpha {\cal H}$, we have
$${\cal H}(\xi^g ) = G^\dagger (g) {\cal H}
(\xi ) G(g),\eqno{(2.13)}$$
leading to
$$
(\partial_\alpha {\cal H }) (\xi^g )=
{\cal D}_{\alpha\beta} (g) G^\dagger (g) 
(\partial_\beta {\cal H})(\xi ) G(g).\eqno{(2.14)}$$
The symbols $\xi$, $\alpha$, $\partial_\alpha$, are natural
generalizations of their continuum counterparts in the previous
subsection. Readers who find this confusing are referred to
Appendix B. 
Ordinary perturbation theory tells us that:
$$\eqalign{
( \partial_\alpha |v>)_\perp (\xi) = &{1\over{{\cal H }(\xi)
-E_0( \xi ) }} |w(\xi )>,~~{\rm where}\cr
|w(\xi )> &\equiv \big [ <v(\xi) |\partial_\alpha 
{\cal H}(\xi ) |v(\xi ) > -\partial_\alpha 
{\cal H} (\xi )\big ] |v(\xi )>.\cr}
\eqno{(2.15)}$$
The above formula is well defined since $<v(\xi) |w (\xi ) >=0$. 
Now, eq. (2.13) implies:
$$|v(\xi^g )> =e^{i\phi_v (\xi , g )} G^\dagger (g) |v(\xi )>.\eqno{(2.16)}$$
The phase factor is arbitrary. Also, $G(g)|v_+> = e^{i\alpha (g)} |v_+>$,
where the phase provides a one dimensional representation of the group
of gauge transformations. 
A simple calculation,
$$\eqalign{&
( \partial_\alpha |v>)_\perp (\xi^g ) =\cr G^\dagger (g)&
{1\over{{\cal H}(\xi ) -E_0 (\xi ) }}
(\xi) G (g) {\cal D}_{\alpha\beta} (g)
\big [  <v(\xi^g )| G^\dagger (g) \partial_\beta {\cal H} (\xi ) 
G(g) |v (\xi^g )> - \cr
G^\dagger (g ) & \partial_\beta {\cal H} (\xi ) G(g) \big ]
|v (\xi^g )> \cr
& = e^{i\phi_v (\xi , g )}{\cal D}_{\alpha\beta} (g) G^\dagger (g) 
[(\partial_\alpha |v>)_\perp ](\xi),\cr}\eqno{(2.17)}$$
shows covariance:
$$
{{<v_+ |[( \partial_\alpha |v >)_\perp ](\xi^g )}\over
{<v_+ | v > (\xi^g )}}=
{\cal D}_{\alpha\beta} (g) 
{{<v_+ |[( \partial_\beta |v >)_\perp ](\xi)}\over
{<v_+ | v> (\xi )}}.\eqno{(2.18)}
$$

Note the important disappearance of the unknown phase due to the cancelation
between the numerator and denominator. This is Fujikawa's view [\fujik]
of the
gauge non-invariance being restricted to the ``fermionic measure'' at work.
To compute $j^{\rm cov}$ in the lattice overlap one does not
need to make a phase choice. Thus, the covariant currents are
defined naturally and gauge covariantly.\footnote{${}^{f_3}$}{This
is particularly useful for QCD applications [\QCDovlap], but a more detailed
discussion would take us too far off track here.} The same goes
for the covariant anomaly: it has no dependence on the phase choice
in the overlap.\footnote{${}^{f_4}$}{Essentially, this is why 
in the original proposal of Kaplan [\kap] it was possible
to compute the anomaly without any apparent ambiguity: 
the outcome turned out to be the covariant anomaly [\nacul].}

In the continuum
we know that for non-singlet non-abelian anomalies $\Delta j$ vanishes
if and only if we have anomaly cancelation, because if there are 
anomalies the consistent and covariant currents can't be equal.
So, anomaly cancelation in this case is equivalent to the vanishing
of $\Delta j$. In the overlap, the main impediments
to arrange for $\Delta j$ to vanish identically 
by deforming the overlap lattice 
Hamiltonians used in the construction of
the states $|v_\pm >$ will be seen to disappear if anomalies cancel. 
The natural and direct definition
of $\Delta j$ and its curl in the overlap
easily extends to the abelian case, unlike
a definition based on (2.10). Just like the covariant current, the 
curl of $\Delta j$ is also
defined in a gauge invariant manner. 

The formula for $\Delta j$ in the overlap is simple:
$$
\Delta j_\alpha = <v|\partial_\alpha |v >\eqno{(2.19)}$$
$\Delta j$ depends on the phase choice for $|v\! >$, but,
the curl of  $\Delta j_\alpha $ ($d\Delta j $) does not; 
if $d\Delta j \ne 0$ there is no way 
a phase choice could eliminate $\Delta j$. 
Whether or not there is a nonzero
curl depends only on the Hamiltonian. It is useful now to recall 
Berry's phase [\berry]. 
Clearly,  $\Delta j$ is nothing but the Berry connection [\simon],
while the curl is Berry's curvature [\berry, \simon]. 

The curl of $\Delta j$ 
was analyzed in [\ball] in the continuum. The curl,
(an abelian field strength over the space of $\xi_\alpha$), in
components ${\cal F}_{\alpha\beta} = \partial_\alpha \Delta j_\beta 
-\partial_\beta \Delta j_\alpha$, is more conveniently manipulated 
after contraction with two arbitrary ``vectors'' $\delta^1_\alpha$
and $\delta^2_\beta$. The 
quantity ${\cal F}_{\alpha\beta} \delta^1_\alpha \delta^2_\beta $ 
is denoted in references [\ball ] by a 
functional $ Z ( \delta_1 A ,\delta_2 A , A) $ 
where $\delta_1 A_\mu (x)$ 
and $\delta_2 A_\nu (y)$ play the role of $ \delta^1_\alpha $ 
and $\delta^2_\beta$. To us the most important 
aspect of the analysis in [\ball]
is that the curvature is finite and apparently  
unambiguous in the continuum regularization 
adopted there which required no gauge breaking even at
intermediary steps. Thus, the curvature is potentially
just as fundamental as the anomaly itself. On the lattice, the
overlap provides a nonperturbative framework to realize the same
situation.

We know that there are
typically two sources to Berry's curvature: One consists of a
``smeared'' collection and the other of ``monopole'' singularities [\berry].
The ``monopole'' singularities cannot
be made to go away by small deformations of the Hamiltonian. But
the smeared component of the source can.

\vskip 1cm
{\bf 3. Two dimensions: Berry's curvature and the abelian anomaly.}
\vskip .5cm
In this section we shall see that anomalies
indeed correspond to  the ``monopoles'' 
identified by Berry, and that anomaly
cancelation corresponds to them canceling each other. The
connection between anomalies and Berry's phase is different from
previous relations described in the literature [\sonoda] and
is specific to the overlap framework, but not to the particular
form of regularization within which the overlap is implemented. 

We are considering a Weyl fermion in two Euclidean dimensions in an
abelian external $U(1)$ gauge field. Space-time is taken to be
a flat torus obtained by identifying the
opposite boundaries of a square. The torus is replaced by a mesh
of small squares and covered by $L^2$ such plaquettes. 

A family of gauge backgrounds is 
chosen to consist of a collection of constant gauge potentials.
Thus, we are concentrating
on a flat torus embedded in the space of gauge orbits. 
There are no exceptional points (in the sense of the introduction)
on this torus, so it is a compact, smooth manifold. Our 
objective is to show that Berry's curvature associated with
the one form $<v|dv>$ on this manifold integrates to a non-zero
value one could associate with degeneracies ``nearby''. The degeneracy
points lie outside the space of $\xi$'s, and assume the role of 
``monopole'' sources. Since we are dealing with the integral of
a locally exact form, its value is quantized, and small deformations
of the Hamiltonian cannot remove the singularity. 
The total strength of the
singularities is $q^2$ where $q$ is 
the integral $U(1)$-charge
of the Weyl fermion, here assumed to be a left mover. For
a right mover we get $- q^2 $. For the integral of the total 
curvature over the torus to vanish the well known anomaly cancelation
condition $\sum_R q_R^2 =\sum_L q_L^2$ must hold.  In that
case, the ``monopoles'' cancel each other out. 

We replace the variables $U_\mu (x)$ by $e^{ih_\mu}$ for each $x$
in eqs. (A.3). By gauge invariance the variables $h_\mu$ are 
periodic with period ${{2\pi}\over L}$ each. 

The matrix $H$ block diagonalizes to two by two
blocks in Fourier space and the ground state of ${\cal H}$ is
obtained by filling the negative energy state corresponding to each
block. We label the momenta and the associated blocks and states
by a two component integral vector $n$, ($n_\mu =0,...,L-1, ~\mu=1,2$). 

For $h_\mu =0$ we have:
$$
p_n = {{2\pi}\over L} {n},~~~~~~~~
H_n =  \pmatrix {{1\over 2} \hat p_n^2 -m & i\bar p^1_n -\bar p^2_n \cr
-i\bar p^1_n -\bar p^2_n & m-{1\over 2} \hat p_n^2\cr}.
\eqno{(3.1)}$$
We use $\bar p_\mu =\sin p_\mu$ and $\hat p_\mu = 2\sin {{p_\mu }\over 2}$. 
For arbitrary $h_\mu$
we simply need to replace every $p_n$ by $p_n +h$.

The curvature we wish to compute is given in second
quantized language by
$$
{\cal F}_{\alpha \beta} = <\partial_\alpha v | \partial_\beta v>-
<\partial_\beta v | \partial_\alpha v>,\eqno{(3.2)}$$
where $\alpha = (\mu , x)$ and $\beta= (\nu , y)$. For our background
we get
$$
\sum_{x,y} {\cal F}_{\alpha \beta} \equiv f_{\mu\nu} =
 <{{\partial v}\over{\partial h_\mu }} 
|{{\partial v}\over{\partial h_\nu }} >-
 <{{\partial v}\over{\partial h_\nu }} 
|{{\partial v}\over{\partial h_\mu }} >.\eqno{(3.3)}$$

The two form $f$, $f=
{1\over 2} f_{\mu\nu}dh_\mu dh_\nu$, 
is taken over the $h$-torus. Define the
numbers ${\tilde f} (m)$ for integral two dimensional
vectors $m$:
$$
{\tilde f} (m) = \left ( {L\over {2\pi}}\right )^2
\int_{|h_\mu|\le {\pi\over L}} e^{-iLh\cdot m} f(h).
\eqno{(3.4)}$$
${\tilde f} (0)$ will be seen to be quantized and
controlled solely by the ``monopoles''; absence
of anomalies is equivalent to the cancelation of
all the ${\tilde f} (0)$ terms among all fermion 
species. Eq. (3.4) can be inverted:
$$
f_{12}(h) = 
\sum_{m \in Z^2} e^{iLh\cdot m} {\tilde f}_m .\eqno{(3.5)}$$
It is straightforward to see that the equation $da=f$
for an unknown one-form $a(h)$ has solutions over the torus 
if and only if ${\tilde f} (0) =0$. The undefined 
part of the
one form $a$ that can be written as $d\Phi$ can 
be eliminated by a phase choice for the ground state;
the rest can only be eliminated by deforming the
matrix $H$. 

Writing out explicitly the Slater determinant wave
function for the 
ground state of ${\cal H}$ in terms of
single particle wave functions one easily derives
$$
f_{\mu\nu} (h) =
\sum_n \big [ 
{{\partial u^\dagger (p_n +h )}\over {\partial h_\mu}}
{{\partial u (p_n +h )}\over {\partial h_\nu}}-
(\mu\leftrightarrow \nu) \big ],\eqno{(3.6)}$$
where $u(p_n )$ is a normalized negative energy 
eigenstate of $H_n$. Clearly, such an object
carries a phase choice, and we wish to make it 
explicit that $f$ does not depend on this phase choice.
For this we need the projector $P_n (h)$ on the
appropriate eigenspace, and an expression for $f$
in terms of $P_n (h)$'s only. 
The (easily proven) required expression can be found in 
[\Avron]:
$$
<\delta_1 u | \delta_2 u > -(1\leftrightarrow 2)=
tr (\delta_2 P P \delta_1 P -(1\leftrightarrow 2) )
,\eqno{(3.7)}$$
where $P=|u><u|$ and $P^2=P$. 
All we need at the moment for proceeding is that
the $H_n (h)$ are two by two hermitian traceless
matrices, and therefore:
$$
P_n (h) = {1\over 2} (1-{\vec w_n }(h) \cdot 
{\vec \sigma}).\eqno{(3.8)}$$
Here, ${\vec \sigma}$ 
is the usual triple of Pauli matrices 
and the real three vectors ${\vec w_n (h)} $ have unit
length. Up to a positive prefactor we have $H_n (h)
\propto {\vec w_n} \cdot{\vec \sigma}$. 
Since $tr(\delta_1 P \delta_2 P) = tr (\delta_2 P \delta_1 P)$, 
we can write:
$$
<\delta_1 u | \delta_2 u > -(1\leftrightarrow 2)=
tr(\delta_2 P (1/2 - P ) \delta_1 P) -
(1\leftrightarrow 2).\eqno{(3.9)}$$
Simple algebra now produces:
$$
<\delta_1 u | \delta_2 u >(p_n +h)  -(1\leftrightarrow 2)=
{i\over 2} {\vec w_n}
\cdot \delta_1 {\vec w_n } 
\times \delta_2 {\vec w_n }
.\eqno{(3.10)}$$
We recognize the appearance of the infinitesimal
area element on the surface of the
sphere ${\vec w_n^2} =1$. 
$$
f_{12} (h) ={i\over 2} \sum_n {\vec w_n}\cdot 
{{\partial {\vec w_n}}\over {\partial h_1}}\times
{{\partial {\vec w_n}}\over {\partial h_2}}.\eqno{(3.11)}$$
The expression above needs to be integrated over
the $h$-torus. For any function $F$ we have:
$$
\int_{|h_\mu |\le {\pi\over L}} d^2 h
\sum_n F(p_n + h ) = \int_{|\theta_\mu |\le \pi} F(\theta ).\eqno{(3.12)}$$
Therefore (see(3.4)), 
$$
{\tilde f } (0) = i{{L^2}\over{2\pi}} N,\eqno{(3.13)}$$
where $N$ is the number of times the torus $|\theta_\mu|\le \pi$ wraps 
around the sphere $w^2=1$
under the map $\theta \longrightarrow {\vec w} $, explicitly
given by:
$$
\pmatrix {{1\over 2} \hat \theta^2
 -m & i\bar \theta^1 -\bar \theta^2  \cr
 -i\bar\theta^1 -\bar\theta^2   & m-{1\over 2} 
\hat \theta^2 \cr}
\equiv E(\theta ) {\vec w}(\theta)\cdot{\vec \sigma}
.\eqno{(3.14)}$$
By definition,
$E(\theta ) \ge 0$ and ${\vec w^2} (\theta ) =1$. 
It is easy to see that the mapping near the south
pole of the ${\vec w}$-sphere is one to one, so $N=1$.

The source of the winding is easily identified if
one considers deforming the parameter $m$ to negative
values. In that case the south pole is never reached
so $N=0$. When $m=0$ there is a degeneracy
at $\theta_\mu =0$. This degeneracy is a Berry ``monopole''. 
When $m$ goes to $-\infty$ the 
torus gets mapped into a single point on the sphere.
Considering the images of the torus as a function of
the parameter $m$ we acquire the picture of a monopole
traveling from the ``outside'' of the torus into its ``inside''
as $m$ is increased from $-\infty$ through zero, and
once it crosses into the ``interior'' the winding number
changes from zero to unity. We conclude that to obtain
the quantity ${\tilde f} (0)$ we only need to survey 
the various zero energy degeneracy points one can induce
by varying also the parameter $m$. These degeneracies
occur in the first quantized Hamiltonian matrix
and in the ground state of the
second quantized Hamiltonian operator simultaneously. 
Only degeneracies at zero energy of $H$ 
play the role of ``monopoles'' for the overlap.

Until now, we dealt with a fermion of unit charge. If
the fermion has charge $q$, $N$ gets replaced 
by $q^2 N$. Actually, one has $q^2$ distinct monopoles,
not one monopole of strength $q^2$. 

To change handedness we know 
that we should replace in the overlap formula
the lowest energy states by the 
highest energy ones. This simply 
amounts to replacing the projectors $P$ by $1-P$,
inducing a sign switch in $N$, and having the expected
effect on the anomaly. 

The calculation showed that, regardless of the phase
choice, i.e. before any gauge non-invariant step
has been taken, we can conclude that the antisymmetric tensor
${\cal F}_{\alpha\beta}$ has a constant piece
given by:
$$
\pm q^2 {i\over{2\pi}} \epsilon_{\mu\nu} \delta_{xy}\equiv \pm q^2 
{i\over{2\pi}} 
{\cal E}_{\alpha\beta}
.\eqno{(3.15)}$$
This equation is consistent with the following continuum expression:
$$
Z (\delta_1 A, \delta_2 A, A) = 
 \pm i{{q^2}\over {2\pi }} \epsilon_{\mu\nu}
\int d^2 x \delta_1 A_\mu (x) \delta_2 A_\nu (x) .\eqno{(3.16)}$$

${\cal E}_{\alpha\beta}$ is a covariant (trivially,
as it is constant), antisymmetric
tensor in $\xi$-space. Such a tensor is available because
of the existence of $\epsilon_{\mu\nu}$ in two dimensional space-time.
Equation $(3.16)$ agrees with [\RDS,\ball]. 

Note that the calculation was done on a finite lattice.
The torus that we used was not in
momentum space, as the latter is discrete. 
Also note that the anomaly is
traced to a quantized integer already on the lattice,
so there is no question about the continuum limit.
This is in general line with the basic philosophy
of the overlap, namely that anomalies should appear as 
phenomena completely divorced from  ultraviolet effects.

\vfill\eject
\vskip 1cm
\centerline{\bf 4. Two dimensions: An example of partial improvement.}
\vskip .5cm
If anomalies do not cancel there is no way to proceed to 
eliminate the curl of $\Delta j$. If this cannot be done,
there is no hope to find a phase choice for the overlap
that would render a vanishing $\Delta j$, and a gauge invariant
chiral determinant. On the other hand, we would like to
be able to arrange for $\Delta j$ to vanish if anomalies do
cancel, because then the regulated determinant would be gauge
invariant, since its derivatives w.r.t. to the gauge fields would
be given by the covariant current while the latter is defined
in a gauge invariant manner and is curl free. 
I don't expect a simple solution for
all possible gauge backgrounds. 
In this section we focus on a subset of backgrounds
over which the curl of the $\Delta j$ corresponding
to an abelian anomaly free two dimensional chiral model is
set to zero by a 
simple adjustment in the regularization.

The model is a favorite of overlap
work [\412]: It contains one periodic 
left mover of charge 2 and four right movers
of charge 1. We know that to consistently quantize the
model on a two torus, one needs to pick the four right movers
to obey the set of four boundary conditions (PP), (PA), (AP) and (AA).
Here, (PA) for example, means periodic in direction 1 and anti-periodic
in direction 2. 

We still restrict our attention to the $h$-torus.
The total curvature will be the algebraic sum of the
curvatures contributed by each one of the five fermions. Per fermion
these contributions have the form of equation (3.6). Generically it
looks like:
$$
f_{\mu\nu} (h) = \epsilon_{\mu\nu} \sum_n \hat f (p_n + h).\eqno{(4.1)}$$

Let us assume that the Hamiltonians for each fermion species
are regularized on similar square lattices. If the size of the lattice
is $L_F$ for fermion $F$ of charge $q_F$ and the argument $h$ in (4.1)
is $h_F$, the combination ${{L_F h_F}\over {q_F}}$ 
must be $F$-independent. 
The function $\hat f$ for each fermion depends only on the form
of the Hamiltonian. Let us take the simple case that all Hamiltonians
are picked of the same form. Let the charge 2 fermion 
live on a lattice of size $L$, and have charge $q=2$. Let the charge
1 fermions all live on lattices of the same size $L_1$. We choose to
work with $h$-variables periodic with period ${{2\pi}\over L}$. 
To implement the boundary conditions we introduce 
variables $n_{PP},~n_{PA},~n_{AP},~n_{AA}$:
$$
n_{PP} \equiv n;~~~n_{PA}\equiv n+(0, 1/2 );~~~n_{AP} 
\equiv n+ (1/2 ,0);~~~n_{AA}\equiv n+(1/2 ,1/2 ).
\eqno{(4.2)}$$
The total curvature is then:

$$\eqalign{
f^{\rm total}_{\mu\nu} (h) =& \epsilon_{\mu\nu} 
\bigg [ \sum_{n;~n_\mu=0,1,...L-1} \hat f ({{2\pi}\over L} n + 2h )-\cr
 &\sum_{n;~n_\mu=0,1,...L_1 -1} \hat f ({{2\pi}\over L_1} n_{PP} 
 + {L\over {L_1}} h )-\cr
 &\sum_{n;~n_\mu=0,1,...L_1 -1} \hat f ({{2\pi}\over L_1} n_{PA}
 + {L\over {L_1}} h )-\cr
 &\sum_{n;~n_\mu=0,1,...L_1 -1} \hat f ({{2\pi}\over L_1} n_{AP} 
 + {L\over {L_1}} h )-\cr
 &\sum_{n;~n_\mu=0,1,...L_1 -1} \hat f ({{2\pi}\over L_1} n_{AA} 
 + {L\over {L_1}} h ) 
\bigg ].\cr }\eqno{(4.3)}$$

Clearly, the choice $L_1 = L/2$ produces
exact cancelation and zero total curvature on the $h$-torus. 
The set of boundary conditions for the charge 1 fermions 
work precisely so that each one of the four monopoles associated
with the charge 2 fermion is individually canceled by a 
monopole associated
with one specific charge 1 fermion. 
The choice of $L_1 =L/2$ was implemented in previous work [\num412],
but its impact on the phase of the overlap was not mentioned.

It is not necessary to have a strictly gauge
invariant chiral determinant for all gauge backgrounds;
one can rely on gauge averaging and the Foerster, Nielsen, Ninomiya 
mechanism instead [\FNN]. 
Nevertheless, 
in the numerical simulations carried out for this model [\num412] 
it was important
to use a definition of the overlap that was close to ``ideal'' 
at least
for the gauge field configurations that carry significant
statistical weight when the lattice coupling becomes large
and continuum is approached. 
Since constant gauge fields are not suppressed
in the continuum limit, getting close to a gauge invariant definition
on the $h$-torus discussed above was not only nice but
actually necessary in practice. 

The reason for the necessity is not
fundamental, but has to do with the inability
of Monte Carlo techniques to deal reliably with complex
measures. The phase of the integrand in the integral estimated 
by the Monte Carlo
procedure was absorbed into the observable. Typically this would lead
to disastrous results since wild fluctuations of the phase are expected, 
so impractical long simulation times would be needed to dig the signal out
from under the noise. (Actually, this may not be possible even in
principle because of roundoff errors reflecting the finite number
of digits used to represent floating point numbers on a computer.)
In the particular model of [\412] the continuum limit
is exactly soluble, and tells us that the chiral determinant is 
actually positive. Thus all the phases we would see are lattice artifacts
and, in principle, would be completely eliminated in an ``ideal'' 
regularization.

In practice, the cancelation of the curvature on the $h$-torus
combined with the BW phase 
choice (see section 6 for a definition) turn out to be sufficient. As 
mentioned above, the FNN mechanism indicates that it is
not necessary to have an ``ideal'' regularization; it is the practical
aspects of Monte Carlo integration that make it necessary to
go some distance towards an ideal regularization.
Of course,
it would be nice to know that there exists, in principle,
a choice of Hamiltonians
and other regularization details (something one could call 
a perfect, or ideal, overlap - see 
section 12, bottom of page 380, in [\npblong])
that provide strict
gauge invariance for anomaly free theories. 
On the other hand, it is also important to know 
(at least on the lattice)  
for sure whether, in principle, any {\it fine} tuning is needed
in order to regularize anomaly free chiral gauge theories.
My feeling, based on the FNN mechanism and available numerical
evidence to date is that no {\it fine} tuning is needed in principle.
At present,
this opinion does not appear to be widely shared among workers in the
field.

\vskip 1cm
\centerline{\bf 5. Four dimensions: Berry's curvature and the abelian anomaly.}
\vskip .5cm
In four dimensions we again consider a charged 
left handed Weyl fermion
interacting with a $U(1)$ gauge field. First, we need to identify
a sub-manifold of $\xi$-space on which it is easy to show that
the curl of $\Delta j$ cannot be made to vanish by small smooth
deformations of the Hamiltonian matrices. 

In section 2 we learned that the relevant obstructions 
can be found by varying $m$
and looking for zero energy degeneracies in the Hamiltonian matrix.
Counting the parameter $m$ as 
one of the dimensions, in addition to $\xi$, the above degeneracies
are co-dimension three points. 
The easiest case would be a two dimensional 
sub-manifold in the space of gauge orbits
on which the Hamiltonian globally breaks up into two by two blocks.
By varying $m$ we can search for ``monopoles'' associated with
any one of the $2\times 2$ blocks. 
We pick two directions, say 1 and 2, and make the corresponding
components of the gauge potentials in those directions constant. 
This generates a torus over which Berry's curvature
could integrate to a non-zero value. For this we need 
a covariant (constant)
antisymmetric tensor with two indices, $\alpha$ and 
$\beta$, as before. On the torus there are no
space-time variables we can use to produce antisymmetry so
we need to use 
the $\epsilon_{\mu\nu\rho\sigma}$ tensor. 
To absorb two of its indices the simplest
is to introduce a constant magnetic field in the 3 and 4 directions.
Therefore, we pick the 3 and 4 components of the gauge potential
independent of $x_1$ and $x_2$, but with the dependence on $x_3$
and $x_4$ chosen so as to generate a constant (and hence
quantized) magnetic field
through all 3-4 plaquettes. These are instanton configurations from
the view point of a 3-4 two dimensional world, and they
have been extensively studied before {[\prev2dinst,\PRL]}. 

In summary, we pick the following family of backgrounds:
$$
\eqalign{
U_3 (x) =& \cases { 1,&  $x_3 \ne L-1$ \cr
e^{-i{{2\pi}\over L} x_4 },& $x_3 = L-1$\cr}\cr
U_4 (x) =& e^{i{{2\pi}\over {L^2}} x_3 }\cr
U_1 (x) =& e^{ih_1 }\cr
U_2 (x) =& e^{ih_2 }\cr}\eqno{(5.1)}$$
We assume for the time being unit fermion charge and that the system
is defined on an $L^4$ torus. The backgrounds are parameterized by
a two dimensional torus consisting of points labeled by $h$. 

We
proceed to describe a basis 
in which the Hamiltonian matrix corresponding to
these backgrounds becomes $2\times 2$ block-diagonal.
Since there is no dependence on $x_1$ and $x_2$ we first go to Fourier
space in these variables. We shall denote the momenta by $p_n$ just
as in the previous sections. These momenta are two dimensional. 
 
Employing notations given in the appendix
(using the lower left corner of the list in (A.4)), 
$H$ is explicitly given by:
$$
H=(B-m)\sigma_3 \otimes {\bf 1} +i W_3 \sigma_2 \otimes \sigma_3 -
i W_4 \sigma_1 \otimes {\bf 1} +i\sigma_2 \otimes \left [ W_1\sigma_1
+W_2\sigma_2 \right ].\eqno{(5.2)}$$
One should view $H$ as a square matrix acting on a space of dimension $4L^4$
where the $L^4$ factor comes from the $L^2$ labels $(x_3 , x_4)$ and
the $L^2$ labels $p_n$. The factor 4 comes from the two two dimensional
spaces made explicit in (5.2). 
$H$ is block diagonal in the $n$ indices and its $4L^2 \times 4L^2$ 
diagonal blocks will be labeled by $H^n$. Also, $B_1 +B_2 = {1\over 2}
\hat p^2$. 

Associated with each $n$-block introduce two new $2L^2\times 2L^2$ 
matrices, $H^n_\pm $:
$$
H^n_\pm = \pm i W_3 \sigma_2 -i W_4 \sigma_1 + 
(B_3 +B_4 - m + {1\over 2} \hat p^2 ) \sigma_3.\eqno{(5.3)} $$
The dependence on $p_n$ and $h$ comes in through
the quantity $\hat p^2$, where $p$ stands for the
combination $p_n +h = {{2\pi}\over L} n +h;~~ 0 \le h_\mu < {{2\pi}\over L}$. 
Any $p$ with $0 \le p_\mu < 2\pi$ uniquely
identifies an $h$ and an $n$. As before, the dependence is really only
on the combination $p_n +h$ and this will allow us to use
equation (3.12). 

The matrices $H^n_\pm$ are two
hermitian $d=2$ Hamiltonians in $d=2$ instanton backgrounds. 
Let us introduce their $2L^2$-dimensional eigenvectors:
$$
H^n_\pm \psi^A_{n\pm}= E^A_{n\pm} \psi^A_{n\pm} .\eqno{(5.4)}$$
Since  $H^n_+ =-\sigma_2  H^n_- \sigma_2$ we 
can choose $\psi^A_{n+}\equiv \psi^A_n$
and $\psi^A_{n-}\equiv \hat\psi^A_n =\sigma_2 \psi^A_n$, 
with $E^A_{n+} = E^A_n$ and $E^A_{n-} = - E^A_n$.

Introduce now
the following $4L^2$-dimensional vectors:
$$
\phi^A_n =\psi^A_n 
\otimes \uparrow ,~~~~\hat\phi^A_n =\hat\psi^A_n \otimes \downarrow
.\eqno{(5.5)}$$
The index $A$ takes $2L^2$ values, the index $n$ takes $L^2$ values
and the set $\{ \phi^A_n,\hat\phi^A_n \}$ constitutes an orthonormal
basis of the $4L^4$ dimensional space $H$ is acting on. 
The two dimensional spinors $\uparrow$ and $\downarrow$ are the $\pm 1$
eigenvectors of the $\sigma_3$ factor in ${\bf 1}\otimes \sigma_3$. 
In view of $W_1$ and $W_2$ being proportional to the unit matrix,
$$\eqalign{
H^n \phi^A_n =&  E^A_n \phi^A_n +(iW_1^n +W_2^n )\hat\phi^A_n\cr
H^n \hat\phi^A_n =& -E^A_n \hat\phi^A_n +(iW_1^n -W_2^n ) 
\phi^A_n .\cr}\eqno{(5.6)}$$
Therefore, $H^n$ has nonzero matrix elements only between states
that carry the same index $A$. For each $A$, $H^n$ reduces to a two by
two matrix $H^n_A$ given by:
$$
H^n_A = \pmatrix{E^A_n & iW_1^n + W_2^n\cr iW_1^n - W_2^n & -E^A_n \cr}
\eqno{(5.7)}$$
Since $\det H^n_A = -(E^A_n)^2 -|W_1^n|^2-|W_2^n|^2$ 
for any $A$ and any $n$, 
we know that there are no exceptional configurations
on our torus, so it is a compact and smooth manifold just as it
was in the section 2.

The dependence on $h$ and $n$ enters as follows:
$$
iW_1^n \pm W_2^n = - \bar p_1 \pm i \bar p_2 ,\eqno{(5.8)}$$
and
$$
E^A_n = g_A ( m - {1\over 2}\hat p^2 ).\eqno{(5.9)}$$
The functions $g_A (\mu )$ are simply the eigenvalues of the $d=2$
Hamiltonian $iW_3 \sigma_2 -i W_4 \sigma_1 + (B_3 +B_4 -\mu)\sigma_3$ 
in an instanton background, viewed as function of a mass
parameter $\mu$ and do not explicitly 
depend on $n$ or $h$. We know [\npblong,\PRL] that
as long as $\mu < 0$, $g_A (\mu)$ is bounded away from zero. Thus, 
for $m<0$ all blocks $H^n_A$
stay non-degenerate. However, from numerical work [\npblong,\PRL],
we know that, for exactly one $A$, $g_A (\mu)$ crosses zero
once as $\mu$ goes from negative to a positive value less than 2.
To get a degeneracy in (5.7) we also need, in addition
to $E^A_n =0$, $\bar p=0$ there. Among the four
values satisfying  $\bar p=0$ 
only $p=0$ also sets $\hat p^2 =0$. 
For this value,
we see now that a variation of $m$ will take us through a degeneracy
and an irremovable contribution to $\Delta j$. For the other 
values, ${1\over 2} \hat p^2 \ge 2$, and the 
argument of $g_A$ stays negative
as long as $m\le 2$; thus potential ``monopoles'' and ``anti-monopoles''
associated with fermion doublers are avoided.

If the fermion has
integral charge $q$ there will be $q$ values of the index $A$ 
for which $g_A (\mu )$ would cross zero. Each such crossing is in the
same direction and the combined effect (at infinite
$L$ one expects the crossings to occur simultaneously in $m$) 
can be viewed
as that of one monopole of strength $q$. The analysis of the previous
sections shows that the two dimensional system will then see $q^2$
such monopoles, so the total contribution goes as $q^3$. 
In four dimensions we can decide that all Weyl fermions are
taken as left handed, and then, to 
cancel anomalies we need $\sum_F q_F^3 =0$, as expected. 
A switch in handedness is equivalent to a switch in the sign of $q$.

It remains to identify the normalization. If we increased the
strength of the constant magnetic field through the 3-4 planes
by an integral factor $l$ the number of crossings and,
consequentially, Berry's curvature would increase $l$-fold. Thus,
the curvature is proportional to the total flux through a 3-4 plane
divided by $2\pi$, its smallest quantum. Putting this together
we arrive, in continuum notation, at
$$\eqalign{
Z(\delta_1 A,\delta_2 A ,A) =&
\pm i{{q^3}\over {4\pi^2 }} \epsilon_{\mu\nu\rho\sigma}
\int d^4 x \delta_1 A_\mu (x) \delta_2 A_\nu (x)
\partial_\rho A_\sigma (x)  =\cr &
\pm i{{q^3}\over {8\pi^2 }} \epsilon_{\mu\nu\rho\sigma}
\int d^4 x \delta_1 A_\mu (x) \delta_2 A_\nu (x)
F_{\rho\sigma}. \cr}\eqno{(5.10)}
$$
This agrees with [\RDS,\ball]. It is also convenient to write the answer
in terms of ordinary forms, 
$$
\eqalign{
\delta_1 A (x) = \delta_1 A_\mu (x) dx_\mu ,~~
\delta_2 A (x) = &\delta_2 A_\mu (x) dx_\mu ,~~ 
F (x) = {1\over 2} F_{\mu\nu}dx_\mu dx_\nu , \cr
\epsilon_{\mu\nu\rho\sigma} dx_1 dx_2 dx_3 dx_4 =&
dx_\mu dx_\nu dx_\rho dx_\sigma ,\cr}\eqno{(5.11)}$$
as:
$$
Z(\delta_1 A,\delta_2 A ,A)=\pm i{{q^3}\over {(2\pi )^2 }} \int 
\delta_1 A (x) \delta_2 A (x) F (x).\eqno{(5.12)}$$

The generalization to $d$ dimensions [\ball] for this abelian case is 
$$
Z(\delta_1 A,\delta_2 A ,A)= \pm i{{q^2} \over
{2\pi ({d\over 2} -1 )!}}
\int 
\delta_1 A (x) \delta_2 A (x)  \left ( {{qF}\over {2\pi  }}
\right )^{{d\over 2}-1} (x).\eqno{(5.13)}$$
The above equation indicates the following interpretation: The torus
we are working on is spanned by constant gauge fields in directions 1
and 2. The factor ${{q^2}\over {2\pi}}$ comes from $q^2$ monopoles. 
The rest of the gauge fields are picked to create constant minimal size
magnetic fields through each one of the planes $(3,4)$, $(5,6)$, etc. 
This explains the appearance 
of the field strength to power ${d\over 2} -1$.
Each field strength comes multiplied
by the charge $q$ and divided by $2\pi$ because this is the basic quantum
of flux. There are ${d\over 2} -1 $ factors in all possible
orders, but there should be
only one, and this is fixed by the prefactor ${1\over{({d\over 2} -1 )! }}$.
Using equation (A.4) and the above analysis in four dimensions 
it should be easy to check how this works out explicitly in $d > 4$. 

The setup of fields we employed is very similar to the one 
used in continuous Minkowski
space by Nielsen and Ninomiya [\NNintuit] 
to provide a simple interpretation for
anomalies. The main difference is in the first
two dimensions: in 
Minkowski space one uses a real constant electric field while here,
in Euclidean space, we used only
constant gauge potentials. This difference stems from anomalies in
Euclidean space having to do exclusively with phases. 

It would be instructive to see how much of the present analysis
can be reproduced in the nonabelian case, where one can employ constant
gauge potentials in all directions. This might also
be useful to large $N$ reduced models. 

\vskip 1cm
\centerline{\bf 6. The Brillouin-Wigner phase choice.}
\vskip .5cm
The Brillouin-Wigner (BW) phase choice is defined for 
gauge configurations with a non-degenerate $|v>$, where,
in addition, $<v_0 |v>\ne 0$. $|v_0>$ is a carefully
chosen state, typically the ground state of ${\cal H}$ for the gauge
background $U_\mu (x) =1$ [\npblong,\num412]. The phase
of the overlap is fixed by requiring $<v_0|v>$ to be
positive. 

Actually, an equally appropriate description of the
phase choice would be as Pancharatnam's phase choice
[\pancha]. In the context of the Poincare representation
of light polarizations, two beams $|A>$ and $|B>$ were
defined by Pancharatnam to be in phase if $<A|B>~>0$.
He observed that if $|B>$ is in phase with $|A>$ and 
$|C>$ with $|B>$ then $|C>$ need not be in phase with
$|A>$. This clearly reflects an underlying non vanishing
curvature. It implies that we cannot arrange for all
$|v>$'s to be in phase with each other. 

Berry [\pancha] showed that if $|A>$ was parallel transported
to $|B>$ using Berry's connection on the Poincare sphere,
along the shortest geodesic, $|A>$ and $|B>$ ended up in
phase. If we found a specific regularization that made
$d\Delta j =0$ for a specific anomaly free theory we 
could arrange for all $|v>$'s to be in phase and the
BW phase condition would also have been satisfied. 

Note the related fact that the BW phase condition can
be defined from an extremum principle, namely,
requiring to maximize $\| |v_0> +|v> \|^2$. In the
generic case the extremum principle would guarantee
compatibility of the BW phase choice with symmetries
in the sense analyzed in detail in [\npblong] on a
case by case basis.

\vskip 1cm
\centerline{\bf 7. Discussion.}
\vskip .5cm
Sections 2 and 4 present probably the simplest ``derivation'' of 
anomalies (including their normalization) in any lattice regularization. 
The geometric insight also produces a clean definition of covariant
currents associated with global or local symmetries acting
on the fermions. The separation of sources for  Berry's curvature
into ``monopoles'' and the rest may provide the first step in 
establishing in principle the existence of an ``ideal'' lattice
overlap regularization for any anomaly free gauge theory, chiral
or not. 

A potential example of an ``ideal'' lattice regularization
might be obtained by replacing $B+\gamma_\mu W_\mu$ in (A.2)
by the recently discussed fermionic actions in [\hasen]
and use the new Hamiltonians in (A.5)
to define the $|v>$-line bundle.
One would also need to exactly map the one dimensional line
in lattice action space followed by the marginally relevant
nonlinear flow associated
with a fermion mass and use the theory with the negative mass sign
(assuming exact parity invariance for the pure gauge part of the action)
and replace $m$ accordingly. At least naively, the above  
suggestion brings the lattice overlap ``closest'' (in the
sense of violation of conformal invariance)
to a continuum overlap, and, if the latter
were really defined outside perturbation theory, it should work
as an ideal regularization as long as the mass
parameter is perceived as infinite on physical scales. 
Of course, more work is needed
to test this suggestion.
 
Unfortunately, the definition of the Wilson-Dirac
operator we would use, as 
given in [\hasen], is implicit, requiring the solution of a
complicated non-linear RG recursion relation on an 
infinite lattice, since arbitrarily
separated fermions eventually become coupled (albeit weakly). 
Only after the solution is obtained can we go to
a finite lattice by factoring translations appropriately.  
It hasn't been proven yet that a unique
solution to the recursion relation 
exists, and even if it does, it remains unclear
whether the solution provides a matrix whose entries
are smoothly dependent on the arbitrary link variables.
For the considerations of this paper, a smooth dependence
of the Hamiltonian matrices on the group valued gauge
fields is absolutely essential.

It might be worthwhile to mention here that our four
dimensional backgrounds, for any value of the mass 
parameter in the ranges we considered, can easily be 
deformed to (abelian) configurations for which the 
associated Wilson-Dirac operators have a pair of exact 
zero modes. The deformation is obtained by introducing
a parameter $\kappa$ in the exponents of $U_{3,4} (x)$ 
in equation (5.1) and reducing $\kappa$ from 1 towards zero. 
It is guaranteed that this kind of special 
configurations will be encountered for some 
$0<\kappa <1$, depending on the mass $m$. These
configurations have what would amount to a singularity
in the continuum, and there is no reason to associate
the fermion zero modes with four dimensional instantons.
This may be of interest to present day numerical QCD
work. 

I feel that the geometric insight presented in this paper
reveals one of the deeper aspects of the overlap formalism. 

\vskip 1cm
\centerline{\bf Acknowledgments.}
\vskip .5cm
This work was supported in part by the DOE under grant \#
DE-FG05-96ER40559. I thank Randjbar-Daemi for helpful 
correspondence
relating to reference [\RDS]. 
\vskip 1cm
\centerline{\bf Appendix A.}
\vskip .5cm
Based on [\npblong], we know that the overlap for a chiral fermions 
in even $d$ dimensions, interacting with gauge field in a compact
group $G_s$, is defined as follows:
On the lattice, the chiral 
determinant is replaced at the regulated level by
the overlap of two fermionic many-body states. 
These are the ground states
of two bilinear Hamiltonians,
$$
{\cal H}^\pm = a^\dagger H^\pm a, \eqno{(A.1)}$$
with all indices suppressed. The matrices $H^\pm$ are obtained from
$$
H(m) = \gamma_{d+1} \left [B -m +\gamma_\mu W_\mu \right ]\eqno{(A.2)}$$
with $H^+ = H(-\infty )$, $H^- = H(m_0 )$ and $0 < m_0 < 2$.
The infinite argument for $H^+$ can be replaced by any finite positive
number, but the equations are somewhat simpler with the above choice
[\412,\num412,\minfty]. The matrices $W_\mu$ and $B=\sum_\mu B_\mu $ 
are given below:
$$
\eqalign{
(W_\mu )_{x i, y j}
& ={1\over 2}
[\delta_{y,x+\hat\mu} U_\mu (x)  -
\delta_{x,y+\hat\mu} U_\mu^\dagger (y) ]^{ij} ,\cr
( B_\mu )_{x i, y j} & = {1\over 2} [2\delta_{xy} {\rm\bf 1}
 - \delta_{y,x+\hat\mu} U_\mu (x)  -
\delta_{x,y+\hat\mu} U_\mu^\dagger (y) ]^{ij}.\cr} \eqno{(A.3)}$$
$x,y$ ($x_\mu =0,1,..,L-1$) are sites on the lattice and $i,j$ 
are color indices. The matrices $U_\mu (x)$ are $G_s$
link variables on the lattice, associated with direction $\mu$
and site $x$. The $\gamma_\mu$ are Euclidean
Dirac matrices in $d$ dimensions. We choose the following 
chiral basis, with alternating real symmetric and imaginary antisymmetric,
matrices: 
$$\eqalign{
\gamma_1~~ =& \sigma_1 \otimes \sigma_1 \otimes \sigma_1 .... \otimes 
\sigma_1 \otimes \sigma_1 \otimes \sigma_1 \cr
\gamma_2~~ =& \sigma_1 \otimes \sigma_1 \otimes \sigma_1 .... \otimes 
\sigma_1 \otimes \sigma_1 \otimes \sigma_2 \cr
\gamma_3~~ =& \sigma_1 \otimes \sigma_1 \otimes \sigma_1 .... \otimes 
\sigma_1 \otimes \sigma_1 \otimes \sigma_3 \cr
\gamma_4~~ =& \sigma_1 \otimes \sigma_1 \otimes \sigma_1 .... \otimes 
\sigma_1 \otimes \sigma_2 \otimes {\bf 1~} \cr
\gamma_5~~ =& \sigma_1 \otimes \sigma_1 \otimes \sigma_1 .... \otimes 
\sigma_1 \otimes \sigma_3 \otimes {\bf 1~} \cr
\gamma_6~~ =& \sigma_1 \otimes \sigma_1 \otimes \sigma_1 .... \otimes 
\sigma_2 \otimes {\bf 1~} \otimes {\bf 1~} \cr
\gamma_7~~ =& \sigma_1 \otimes \sigma_1 \otimes \sigma_1 .... \otimes 
\sigma_3 \otimes {\bf 1~} \otimes {\bf 1~} \cr
..........&...............................................\cr
\gamma_{d-3} =& \sigma_1 \otimes \sigma_1 \otimes \sigma_3 .... \otimes 
{\bf 1~} \otimes {\bf 1~} \otimes {\bf 1~} \cr
\gamma_{d-2} =& \sigma_1 \otimes \sigma_2 \otimes {\bf 1~} .... \otimes 
{\bf 1~} \otimes {\bf 1~} \otimes {\bf 1~} \cr
\gamma_{d-1} =& \sigma_1 \otimes \sigma_3 \otimes {\bf 1~} .... \otimes 
{\bf 1~} \otimes {\bf 1~} \otimes {\bf 1~} \cr
\gamma_{d}~~ =& \sigma_2 \otimes {\bf 1~} \otimes {\bf 1~} .... \otimes 
{\bf 1~} \otimes {\bf 1~} \otimes {\bf 1~} \cr
\gamma_{d+1}=& \sigma_3 \otimes {\bf 1~} \otimes {\bf 1~} ....\otimes {\bf 1~}
\otimes {\bf 1~}\otimes {\bf 1~} =
(-i)^{d/2} \gamma_1 \gamma_2 ...\gamma_d 
}\eqno{(A.4)}$$
Each of the first $d$ rows has $d/2$ two by two factors. 
Two inequivalent representations of the $2^{d/2 -1}$
Euclidean Weyl matrices can be obtained by replacing
in $\gamma_\mu$  the first ``$\sigma_1 \otimes$'' by 1 for $\mu =1,..,d-1$ 
and ``$\sigma_2 \otimes$'' by $\mp i$ for $\mu =d$.

The overlap $O$, is given by
$$
O=<v_+ | v_- >,~~~~~~~~~~~
{\cal H}^\pm |v_\pm > = E_{\rm min}^\pm |v_\pm>.
\eqno{(A.5)}$$

$O$ is the
regularized expression for $\det D(\xi )$. 
$E^\pm_{\rm min}$ denote minimal energies which
define the associated states, assuming no degeneracies. 
The 
state $<v_+|$ becomes trivial with the choice  $H^+ = H(-\infty )$.
Choosing exactly one vector in ${\cal V}$ from the set
$\{ e^{i\Phi} |v_- >, 0< \Phi\le 2\pi\}$
for each gauge background, amounts to a ``phase choice'' for $O$.
\vskip 1cm
\centerline{\bf Appendix B.}
\vskip .5cm
The main
purpose of this appendix is to give a detailed derivation of (2.14).

On the lattice $\alpha$ replaces $(\mu,x,ij)$ where, for definiteness
take $G_s = SU(n)$ with $i,j=1,...n$. $\xi$ labels a point in
the space $\prod_{\mu ,x} G_s$. One value of $\xi$ contains a complete
description of the gauge background, namely $n^2-1$ real numbers
per link. 

Let us denote the fermionic operator $\partial_\alpha {\cal H}(\xi)$ by
$R_\alpha (\xi )$. It is defined by:
$$
{\cal H} (\xi +\delta\xi ) -{\cal H} (\xi ) = 
R_\alpha (\xi )\delta \xi_\alpha,
\eqno{(B.1)}$$
where $\delta\xi_\alpha$ is an infinitesimal change in the gauge background,
equivalent to $\delta U_\mu^{ij} (x)$. Define the quantities
$(\delta U^g )_\mu^{ij} (x)$ for a finite lattice gauge transformation
$g^{ij} (x)$:
$$
(\delta U^g )_\mu^{ij} (x) = g^{\dagger ~ik} (x+\mu )
\delta U_\mu^{kl} (x) g^{lj} (x).\eqno{(B.2)}$$
The $\delta U_\mu^{kl} (x)$ are restricted to a $n^2-1$ linear linear space
per link by the linear requirement 
$$
U^\dagger_\mu \delta U_\mu +\delta U^\dagger_\mu U_\mu =0. \eqno{(B.3)}$$
(B.2) ensures 
$U^{g~\dagger}_\mu \delta U^g_\mu +\delta U^{g~\dagger}_\mu U^g_\mu =0$
if (B.3) holds. Equation (B.2) can be rewritten as:
$$
(\delta\xi^g)_\beta = ( {\cal D}^{-1} (g) )_{\alpha\beta} \delta\xi_\alpha .
\eqno{(B.4)}$$
This is the lattice equivalent to the variation of (2.1). 
Now,
$$
{\cal H}((\xi+\delta\xi)^g ) - {\cal H} (\xi^g ) =
{\cal H}(\xi^g +\delta\xi^g ) - {\cal H} (\xi^g )=
 R_\alpha (\xi^g ) (\delta\xi^g )_\alpha = R_\alpha (\xi^g ) 
({\cal D}^{-1} (g) )_{\beta\alpha} \delta\xi_\beta .\eqno{(B.5)}$$
On the other hand,
$$
{\cal H}((\xi+\delta\xi)^g ) - {\cal H} (\xi^g ) = G^\dagger (g)
[{\cal H}(\xi^g +\delta\xi^g ) - {\cal H} (\xi^g )] G^g =
G^\dagger (g) R_\beta (\xi ) G(g) \delta\xi_\beta .
\eqno{(B.6)}$$

Comparing (B.5) and (B.6) gives, on account of the arbitrariness of
$\delta \xi_\alpha$,
$$
R_\alpha (\xi^g ) = ({\cal D} (g))_{\alpha\beta} G^\dagger (g)
R_\beta (\xi ) G(g).\eqno{(B.7)}$$
This is equation (2.14).

\vskip 1cm
\centerline{\bf References.}
\vskip .5cm

\item{[\gaumeginsp]} L. Alvarez-Gaume, P. Ginsparg,
Nucl. Phys. B 243 (1984) 449.
\item{[\plbfirst]} R. Narayanan, H.  
Neuberger, Phys. Lett. B 302 (1993) 62.
\item{[\npblong]} R. Narayanan, H. 
Neuberger, Nucl. Phys. B 443 (1995) 305.
\item{[\daemi]} S. Randjbar-Daemi and J. Strathdee, 
Phys. Lett. B348 (1995) 543; Nucl. Phys. B443 (1995) 386; 
 Nucl. Phys. B461 (1996) 305;
Nucl. Phys. B466 (1996) 335. 
\item{[\RDS]}  S. Randjbar-Daemi and J. Strathdee, 
Phys. Lett. B402 (1997) 134.
\item{[\phillips]} A. Phillips, Annals of Physics 161 (1985) 399.
\item{[\berry]} M. V. Berry, Proc. R. Lond. A392 (1984) 45.
\item{[\BZ]} W. Bardeen, B. Zumino, Nucl. Phys. B244 
(1984) 421. 
\item{[\simon]} B. Simon, Phys. Rev. Lett. 51 (1983) 2167.
\item{[\ball]} R. D. Ball, Phys. Rep. 182 (1989) 1.
\item{[\npbold]} R. Narayanan, H. Neuberger, Nucl. Phys. B412 (1994) 574.
\item{[\fujik]} K. Fujikawa, Phys. Rev. Lett. 42 (1979) 1195.
\item{[\QCDovlap]} H. Neuberger, Phys. Lett. B417 (1998) 141; 
hep-lat/9801031; Phys. Rev. D57 (1998) 5417.
\item{[\kap]} D. B. Kaplan, Phys. Lett. B288 (1992) 342.
\item{[\nacul]} S. G. Naculich, Nucl. Phys. B296 (1988) 837.
\item{[\sonoda]} H. Sonoda, Nucl. Phys. B266 (1986) 410;  
A. J. Niemi, G. W. Semenoff, Phys. Rev. Lett. 55 (1985) 927.
\item{[\Avron]} J. E. Avron, R. Seiler, B. Simon, Phys. Rev. 
Lett. 51, (1983) 51. 
\item{[\412]} R. Narayanan, H. Neuberger, Phys. Lett. B 393 (1997) 360; 
Y. Kikukawa, R. Narayanan, H. Neuberger, Phys. Lett. B 399 (1997) 105.
\item{[\num412]} Y. Kikukawa, R. Narayanan, H. Neuberger, 
Phys. Rev. D57 (1998) 1233.
\item{[\FNN]} D. Foerster, H. B. Nielsen, M. Ninomiya, 
Phys. Lett. B 94 (1980) 135.
\item{[\PRL]} R. Narayanan, H. Neuberger, Phys. Rev. Lett. 
71 (1993) 3251.
\item{[\NNintuit]} H. B. Nielsen, M. Ninomiya, NBI-HE-91-08, (1991),
published in ``Trieste conference on topological methods in quantum
field theories'', pages 149-171, (World Scientific, 1991, editor: W. Nahm). 
\item{[\pancha]} S. Pancharatnam, The Proceedings of the
Indian Academy of Sciences, Vol. XLIV, No. 5, Sec. A, (1956);
M. V. Berry, Journal of Modern Optics, 34, (1987) 1401.
\item{[\hasen]} P. Hasenfratz, V. Laliena, F. Niedermayer,
hep-lat/9801021.
\item{[\minfty]} D. Boyanovsky, E. Dagotto, E. Fradkin, 
Nucl. Phys. B 285 (1987) 340; Y. Shamir, Nucl. Phys. B406 (1993) 90.

\vfill
\eject
\end